\documentclass [a4paper,12pt] {article}

\usepackage{amsmath}
\usepackage[dvips]{graphicx}
\usepackage{subfigure}
\usepackage[section]{placeins}

\newcommand{\D}{\ensuremath{\mathrm{d}}}

\begin {document}


\title {\textbf{An asymmetric exclusion model with overtaking \\
\Large{a numerical and simulation study}}}
\author {A.~Tonddast~Navaei$\mathrm{^{I, IV}}$
 \and V.~Karimipour$\mathrm{^{I *}}$ \and M.~R.~Ejtehadi$\mathrm{^{II,
III}}$ }

\maketitle

\begin{center}
\footnotesize{ $\mathrm{^{I}}$Department of Physics\\ Sharif
University of Technology\\ P. O. Box 11365-9161\\ Tehran, Iran

$\mathrm{^{II}}$Institute for studies in theoretical Physics and
Mathematics\\ P. O. Box 19395-5531\\ Tehran, Iran

$\mathrm{^{III}}$Max-Planck-Institute for Polymer Research
\\Theory Group \\P. O. Box 3148 D-55021\\ Mainz, Germany

$\mathrm{^{IV}}$Department of Environmental and Mechanical
Engineering\\ The Open University\\ MK7 6AA\\Milton Keynes,
United Kingdom\\
 }

\scriptsize{$\mathrm{^{*}}$
\it{email~}\tt{vahid@sina.sharif.ac.ir}}
\end{center}
\vskip1cm


\begin {abstract}
An asymmetric exclusion model on an open chain with random rates
for hopping particles, where overtaking is also possible, is
studied numerically and by computer simulation. The phase
structure of the model and the density profiles near the high and
low density coexistence line are obtained. The effect of a site
impurity is also studied.
\end {abstract}

\linespread{1.3}
\section {Introduction}
Over the past few years, a great deal of progress has been
achieved in our understanding of boundary and disordered-induced
phase transitions in driven diffusive systems. Our understanding
of such macroscopic systems, which are far from equilibrium, is
still in its infancy since we need in this case, to study in
detail the stochastic macroscopic dynamics of the system. It is
only in one dimension that various exact and approximate
solutions (See \cite{der, dev} for a review) have helped us to
study the macroscopic stochastic dynamics of model systems, i.e.
by writing the master equations of the underlying Markov process
in suitable coordinates and solving these equations either
exactly or by numerical-simulation techniques. The simplest such
model is the Totally Asymmetric Simple Exclusion Process (TASEP),
first introduced by Spitzer in 1970 \cite{Sp}, as an example of an
interacting stochastic process \cite{Lig1, Lig2, Sph}. This is a
model incorporating a collection of random walkers, hopping with
equal rates to their right empty site on a one-dimensional lattice
and interacting via exclusion. There is now a rather extensive
literature on ASEP and its various generalizations (see \cite{der,
dev} and references therein).

Among the generalizations, we mention one which is relevant to the
present paper, namely the introduction of disorder in the hopping
rates of particles \cite{Ev1, Ev2, K1, K2, K3, BB}. In
\cite{Ev1, Ev2}, a closed system is considered, there is no
injection to or extraction of particles out of the system. Each
particle $\mu$ , has a hopping rate $p_{\mu}$, taken from a random
distribution, the particles keep their initial order and no
overtaking is possible. The bulk density of particles $\rho$, is
an important control parameter, and it is shown that for
$\rho>\rho_{c}$, where $\rho_{c}$ is a critical density, a
segregated phase appears behind the slowest particle on the ring,
a phenomenon shown \cite{Ev1} to be similar to Bose condensation
in equilibrium statistical physics.

The same process has been simulated \cite{BB} on an open chain
where each particle enters the left end of the system with rate
$\alpha$ and leaves the right end with rate $\beta$, and the
intrinsic hopping probability $p_{\mu}\in[c,1]$ of a particle
$\mu$, is drawn from the distribution
\begin{equation}
f(p) =\frac{n+1}{(1-c)^{n+1}}\cdot(p-c)^{n}
\end{equation}
Using mostly the values $n=1$ and  $c=\frac{1}{2}$, it has been
shown in \cite{BB} that the generic form of the phase diagram of
the single species (pure) ASEP, prevails also in this case; that
is depending on the value of alpha and beta, three phases develop
in the system, namely low-density, high-density and
maximal-current phases.Compared to the pure case, there is only a
slight difference in the shape of the coexistence line between
the low and high density phases. The small effect of disorder on
the phase diagram of the pure ASEP has been attributed \cite{BB}
to insufficient time for density inhomogeneities (platoons) to
develop up to the scale of system size. This in turn has been
attributed to the average passing time of particles ($\tau \sim$
system size $L$) and the approximate rate of growth of platoon
sizes ($\xi(t) \sim t^{\frac{n+1}{n+2}}$) which is small compared
with $L$, when $L \rightarrow \infty$, for any $n$.

The above results all apply to a model in which particles can not
overtake each other. The appearance of high density phase and
traffic jams are quite expected in such a model. One can however
introduce models in which particles can overtake each other
\cite{K1}. It's natural to expect that fast particles have a
possibility to overtake slow ones, and one may ask to what extend
this new element may affect the nature of steady state and the
phase structure of the system. The model of \cite{K1} admits an
exact solution \cite{K3} by way of matrix product ansatz , where
the phases can be determined, although the density profiles are
still very difficult, if not impossible, to obtain analytically in
this model. Also it is difficult to obtain any analytical result
concerning other effects like the effect of impurities or
blockage.

In this paper we study the model of \cite{K1} by computer
simulations, and will obtain the phase structure of the model and
density profile near the high-density/low-density coexistence
line, and compare our results with those of \cite{BB}. We also
study the effect of a fixed blockage in a simple two-species
model.


\section {The model}
The model which has been first introduced in \cite{K1} and shown
to be exactly solvable via the matrix product ansatz, refers to a
process in continuous time, in which each particle of $type~\mu$,
has an intrinsic hopping rate $v_{\mu}$ to its right empty site,
as in the models considered in \cite{Ev1, Ev2, BB}, but when this
particle encounters a site already occupied by a particle of
$type~\mu'$ (with $v_{\mu'}<v_{\mu}$), the two particles exchange
their sites with rate $v_{\mu}-v_{\mu'}$, as if the faster
particle stochastically overtakes the slower one (Figure 1).

\begin{figure}[htb]
\begin{center}
\includegraphics[scale=0.95,viewport=0 110 445 250,clip]{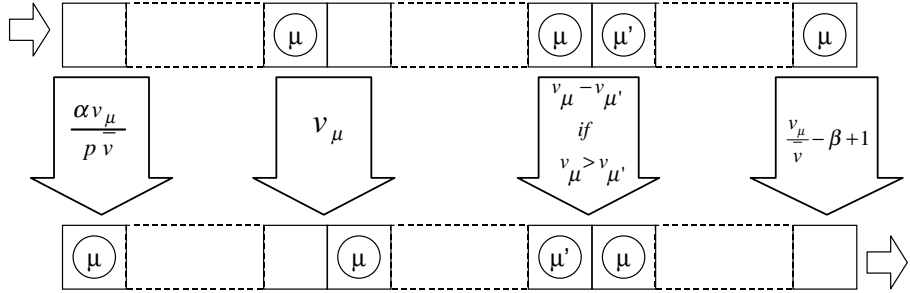}
\end{center}
\caption{The elementary processes.}
\end{figure}

The need for exact solvability of the model, enforces the
injection and extraction rates of particle $\mu$ to be fixed as
\begin{equation}
\alpha_{\mu}=\frac{\alpha v_{\mu}}{p \overline{v}}
\end{equation} and
\begin{equation}
\beta_{\mu}=\frac{v_{\mu}}{\overline{v}}+\beta-1
\end{equation}
where $p$ is the number of type
of particles and
\begin{equation}
\overline{v}:=\frac{1}{p}\sum_{\mu=1}^{p}v_{\mu} .
\end{equation}
It is assumed that there are $p$ type of particles with
hopping rates $v_{1}\leq v_{2}\leq v_{3}\leq \cdots \leq v_{p}$.

Here $\alpha$ and $\beta$ are representing the \emph{total}
injection and \emph{average} extraction rates and
$\beta>1-\frac{v_{1}}{\overline{v}}$, where $v_{1}$ is the slowest
hopping rate. Note that for each particle $\mu$ , $v_{\mu}$ is
its velocity in an otherwise empty lattice, averaged over many
realizations of the process. Thus, hereafter, we use the words
\emph{intrinsic hopping rate} and \emph{intrinsic velocity}
interchangeably. The model can be formulated for continuous
distribution of intrinsic hopping rates which we denote by
$\sigma(v)$. In this case the rate of injection of particles of
intrinsic velocity between $v$ and $v+\D v$ is given by
$\alpha\sigma(v) \frac{v}{\overline{v}} \D v$, where the
extraction rate is given by $(\frac{v}{\overline{v}}+\beta-1) \D
v$. By exactly solving this model \cite{K3}, using the matrix
product ansatz, the phase diagram of this model was determined.
It was shown that besides $\alpha$ and $\beta$, the phase diagram
depends crucially on a characteristic of the distribution
$\sigma(v)$, denoted by $l[\sigma]$ and given by

\begin{equation}
l[\sigma]=\frac{\bar{v}^2}{v_1^2}~-\langle\frac{v\bar{v}}{(v-v_1)^2}\rangle
\end{equation}

where $v_{1}$ is the slowest velocity and the average is taken
with respect to the distribution $\sigma(v)$.

If $\sigma(v)$ is such that $l[\sigma]<0$, then the phase diagram
is similar to the pure ASEP, although the shape of coexistence
line and the position of the triple-point do have essential
dependence on $\sigma$. However when $l[\sigma]>0$, then there is
no high-density phase in the diagram (Figures 2a and 2b).
\begin{figure}[htb]
\begin{center}
\subfigure[$l<0$]{\includegraphics[scale=0.6,viewport=85
510 362 758,clip]{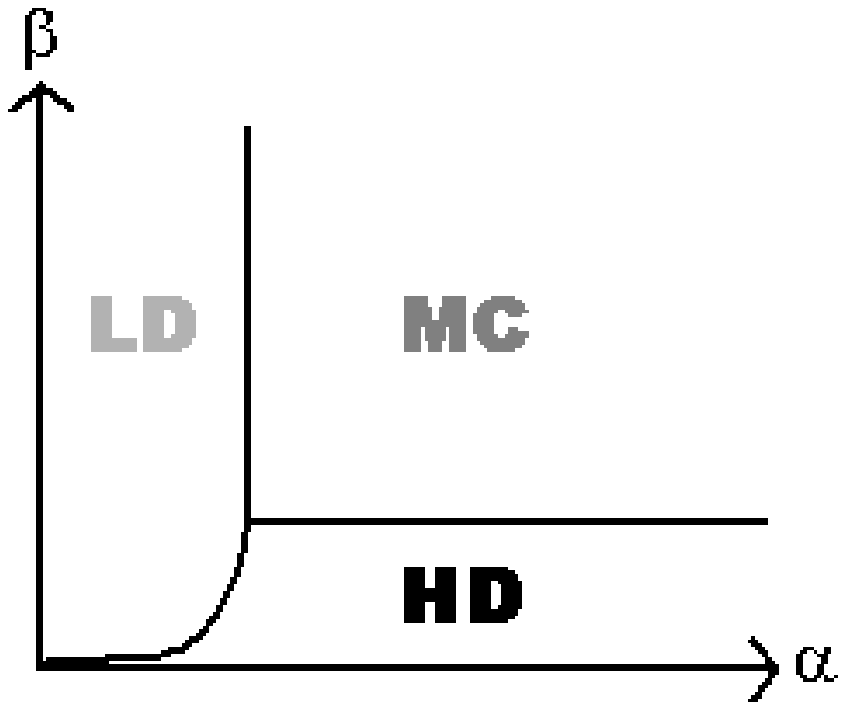}}%
\hspace{\subfigtopskip}%
\hspace{\subfigbottomskip}
\subfigure[$l>0$]{\includegraphics[scale=0.6,viewport=85 510 362
758,clip]{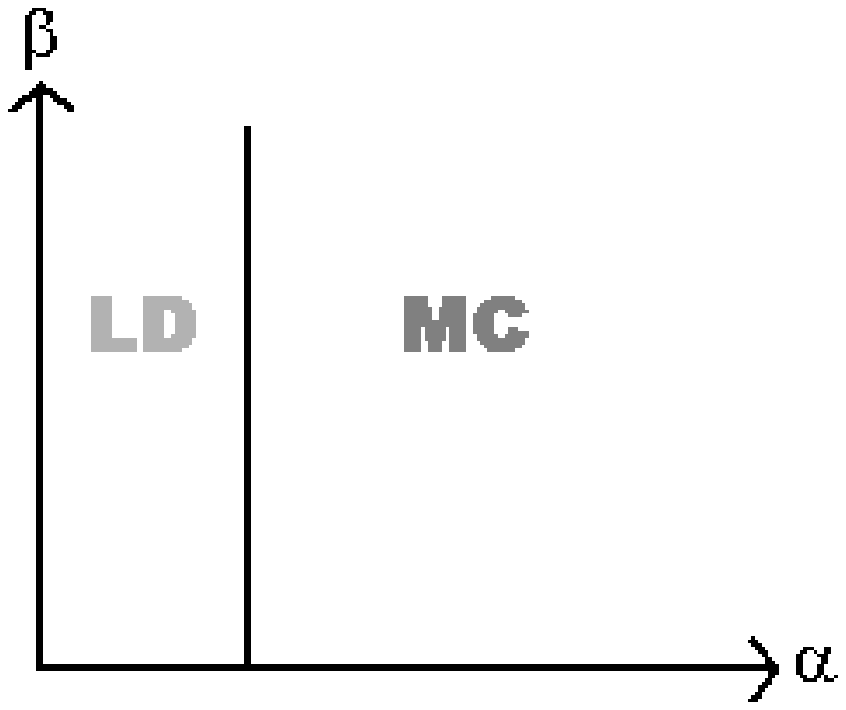}}
\end{center}
\begin{center}
\caption{Typical phase diagrams for two values of $l$.\newline
\scriptsize \sf LD \it low density \sf HD \it high density \sf MC
\it maximal current}
\end{center}
\end{figure}
Specifically for distribution of the $type (1)$, the size and in
fact the existence or non-existence of the high-density phase
depends crucially on the parameters $c$ and $n$\footnote{The
exact criteria is given in the main text.}. This means that if
the chance of entrance of slow particles into the system (cars
into a highway) is sufficiently low, and if there is a
possibility for overtaking slow cars, there will not be a
high-density phase (a traffic jam), regardless of the value of
$\alpha$ and $\beta$. Note that in any case $\beta$ is restricted
by $\beta>1-\frac{v_{1}}{\overline v}$, so that the extraction
rates of all types of particles are positive. In case
$1-\frac{v_{2}}{\overline v}<\beta\leq1-\frac{v_{1}}{\overline
v}$, the steady state will be trivial and the system will
eventually be filled with particles of $type~1$, since in this
case all types of particles except $type~1$ are extracted from
the system. In any other case (e.g. $1-\frac{v_{3}}{\overline v
}<\beta\leq1-\frac{v_{2}}{\overline v}$), the steady state depends
on the initial conditions, since neither particles of $type~1$,
nor $type~2$, are extracted and the steady state depends on the
order of injection of these particles.

If to each site $k$, we assign a random variable $\tau^{\mu}(k)$
which takes the value $1$ only if this site is occupied by a
particle of type $\mu$, then the average density of particle of
type $\mu$ at site $k$, denoted by $\rho^{\mu}(k)$ is
\begin{equation}
\rho^{\mu}(k)= ~ \langle\tau^{\mu}(k)\rangle
\end{equation}
where $\langle~\rangle$ means the long time average, in the steady
state. Note that in accordance with \cite{BB} and in contrast to
the periodic lattice \cite{Ev1, Ev2}, a separate disorder average
is not necessary, since new particles are constantly injected into
the system. Two other quantities are of interest, defined as:
\begin {equation}
\rho^{\mu}:=\frac{1}{L}\sum_{k=1}^{L}\langle\tau^{\mu}(k)\rangle
\end{equation}
\begin{equation}
\rho(k)=\frac{1}{p}\sum_{\mu=1}^{p}\langle\tau^{\mu}(k)\rangle
\end{equation}
which are respectively the density of particles of $type~\mu$,
average over the lattice sites; and the total local density of
particles, irrespective of their type. One can also define the
current of $type~\mu$ of particles, $J^{\mu}$, which in the steady
state is independent of site and in view of the definition of the
process is given by \cite{K1}
\begin{equation}
J^{\mu}= v_\mu \langle\tau_{k-1}^\mu e_k\rangle +
\sum_{\mu'<\mu}(v_\mu -
v_{\mu'})\langle\tau_{k-1}^\mu\tau_k^{\mu'}\rangle -
\sum_{\mu'>\mu}(v_{\mu'}-v_\mu)\langle\tau_{k-1}^{\mu'}\tau_k^\mu\rangle
\end{equation} where
$e_k := 1 - \sum_{\mu=1}^p \tau_k^\mu$ . Note that $\langle e_k
\rangle$ stands for the probability of site $k$ being empty.

 The independence of this current of the site $k$, and its
equality to the input and output current of particles of
$type~\mu$, given by
\begin{equation}
J_{\scriptscriptstyle in}^{\mu}=\frac{\alpha v_{\mu}}{p \overline
v }\langle e_1\rangle
\end{equation}
\begin{equation}
J_{\scriptscriptstyle out}^{\mu}= (\beta+ \frac{v_\mu}{\overline
v}-1)\langle\tau_L^\mu\rangle
\end{equation}
gives us a set of equations, which in the mean field
approximation\\
$\langle\tau^{\mu}(k)\tau^{\nu}(k+1)\rangle\simeq\langle\tau^{\mu}(k)\rangle\langle\tau^{\nu}(k+1)\rangle$
enables all the local densities $\rho^{\mu}(k)$ and hence all the
currents $J^{\mu}$, to be evaluated numerically.

We also study the effect of a site impurity at a given site, say
$k$, by multiplying all the hopping rates by $\lambda~
(\lambda<1)$, when computing the currents of particles on the
links $(k-1,k)$ and $(k,k+1)$. The densities and the current can
again be computed for this case, for various $\lambda$'s and
$k$'s.


\section {Numerical solutions and simulation results}
We consider a lattice of $200$ lattice sites and simulate the
following process with random updating on this lattice:
\begin{description}
\item[a)] A \emph{two} species process, where the hopping rates
are taken to have the distinct values $v_1$ and $v_2$, with equal
probabilities.
\item[b)] a multi-species process, where the hopping rates are
taken from the distribution
\begin{equation}
\sigma(v) = \frac{n+1}{(1-c)^{n+1}}(v-c)^n ~~~,~~~ c\leq v\leq 1
\end{equation}
\item[c)] a \emph{two} species process, where there is a fixed
impurity on one of the sites of the lattice. When the particles
reach this site, they slow down their hopping rate by a factor
$\lambda$. The position of the impurity is denoted by $k$, where
$0\leq k \leq L$.
\end{description}

In cases \textbf{a} and \textbf{c}, we complement our simulations
by a mean field solution obtained by numerically solving the
equations $J_{\scriptscriptstyle
in}^{\mu}=\cdots=J_{(k)}^{\mu}=\cdots=J_{\scriptscriptstyle
out}^{\mu}$.

\subsection*{Results}
\subsection*{a) \begin{normalsize}A two species process\end{normalsize}}

Fixing the values of $v_1$ and $v_2$, we obtain the bulk density
$\rho$, for different values of $\beta$, as function of the
injection rate $\alpha$.

The results are shown in Figures 3a and 3b, for the two different
choices of hopping rates.

\begin{figure}[htb]
\begin{center}
\subfigure[$v_1=0.5,
v_2=1.5$]{\includegraphics[scale=0.55,viewport=0 0 500
390,clip]{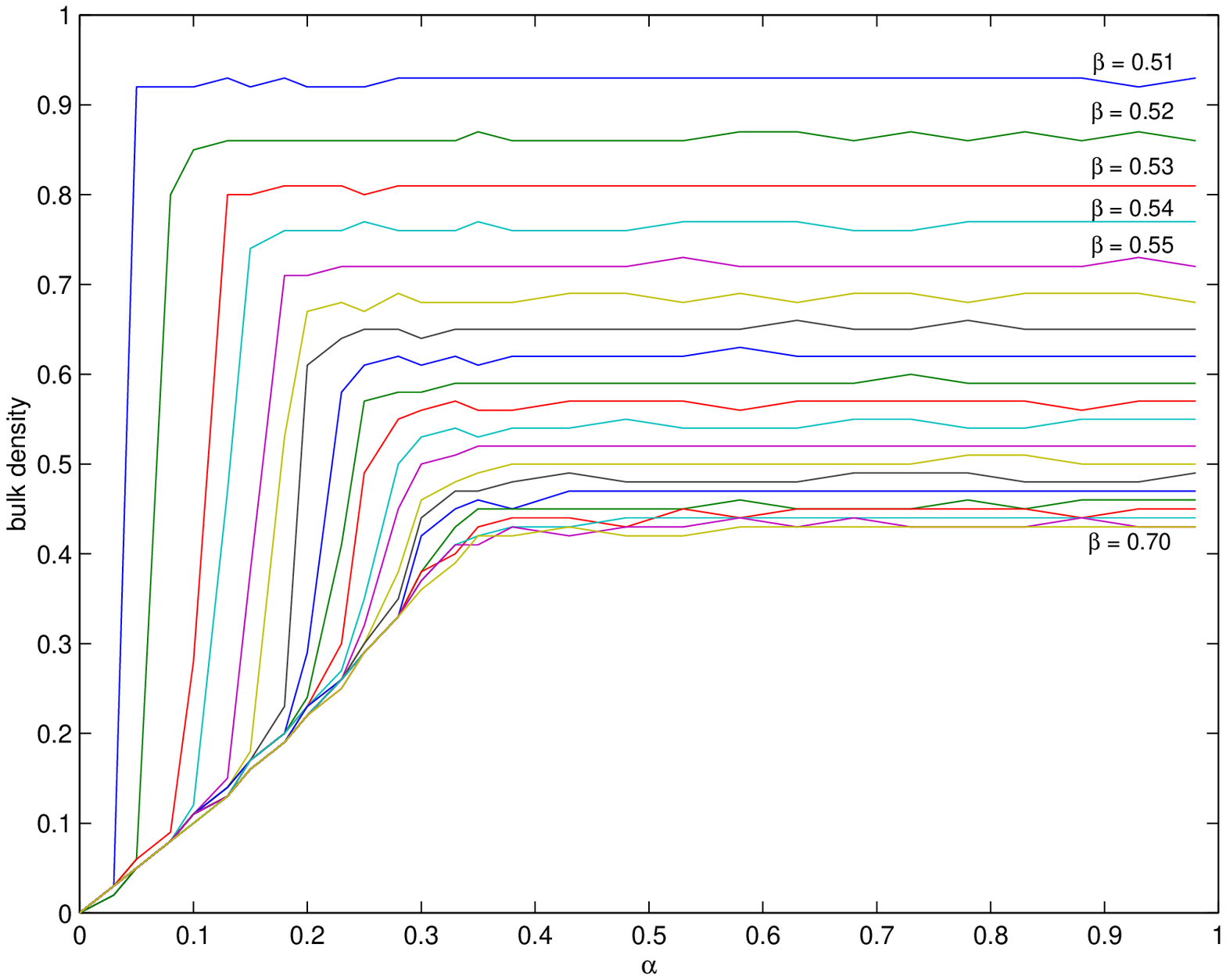}}%
\hspace{\subfigtopskip}%
\hspace{\subfigbottomskip} \subfigure[$v_1=0.8,
v_2=1.2$]{\includegraphics[scale=0.55,viewport=0 0 500
390,clip]{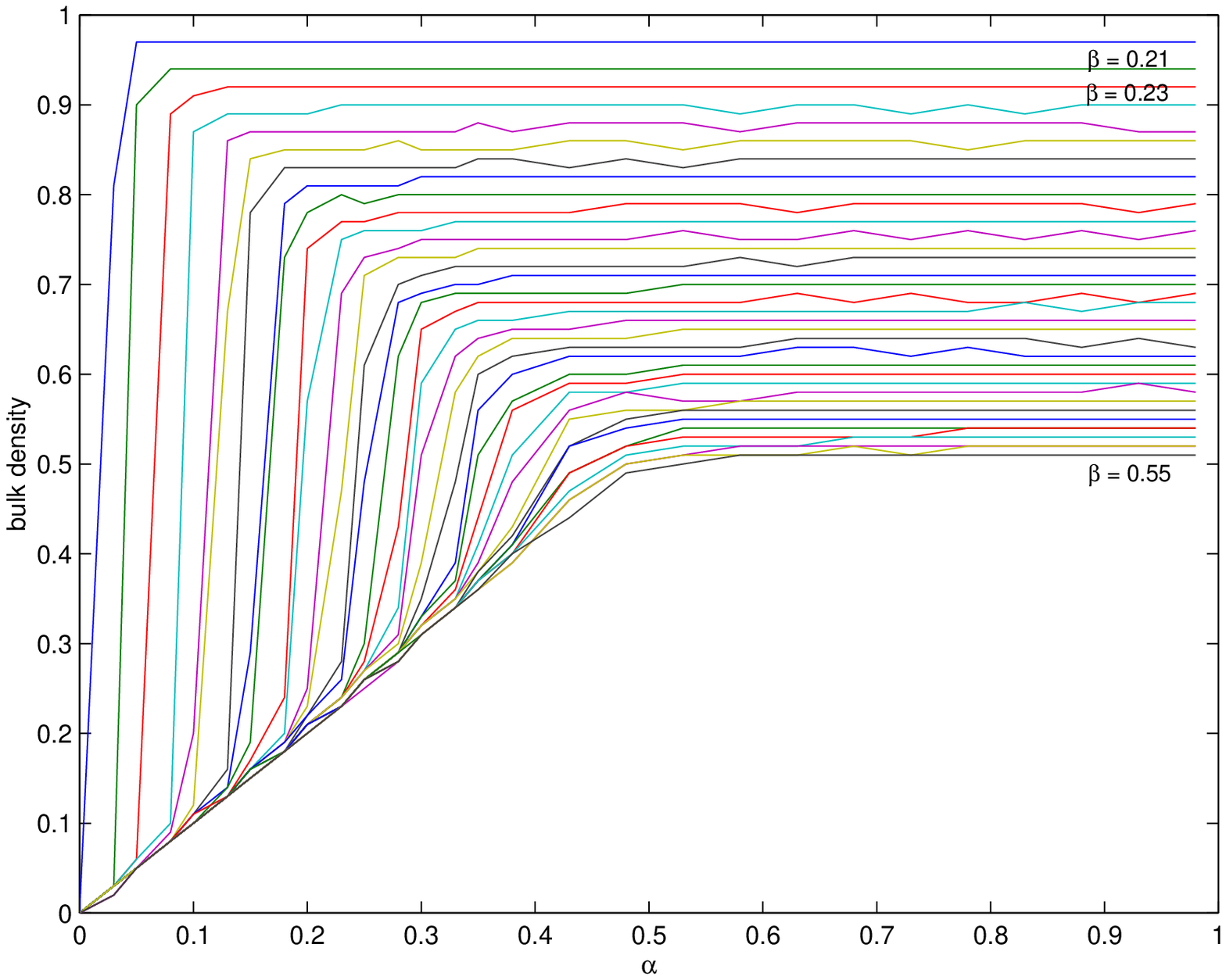}}
\end{center}
\caption{The bulk density as function of $\alpha$ for various
values of $\beta$, in the \emph{two} species model.}
\end{figure}

It is seen that for every $\beta$ , there is a jump discontinuity
in the bulk density at a critical value of
$\alpha_c=\alpha_c(\beta)$. The value of the jump vanishes at a
certain point $(\alpha_c,\beta_c)$, which makes the triple point
in the phase diagram. The corresponding phase diagrams are shown
in Figure 4.

\begin{figure}[h]
\begin{center}
\includegraphics[viewport=0 0 500
390, scale=0.75, clip]{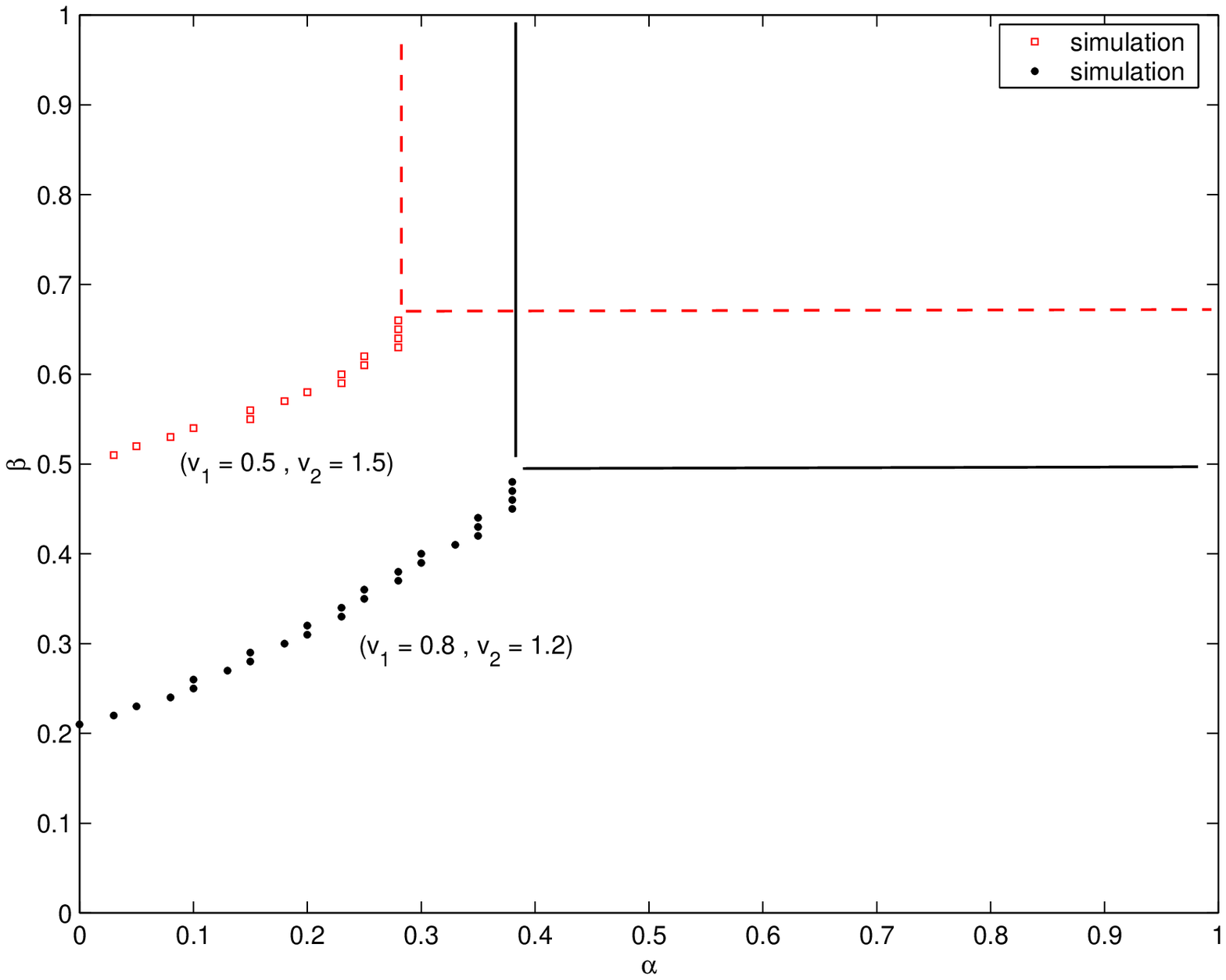} \caption{The phase diagram for
problem \textbf{a}.}
\end{center}
\end{figure}

\subsection*{b) \begin{normalsize}A multi species process\end{normalsize}}
In this case we consider a continuous distribution of hopping
rates. For the sake of comparison with \cite{BB}, we take the
distribution (12).

For this distribution $l[\sigma]$ is calculated to be \cite{K2}:

\begin{equation}
l[\sigma]= \bigg\{ \begin{array}{lll} -\infty & \tt{if} & 0\leq n
\leq1\\
(\frac{\bar{v}}{c})^2-(n+1)\bar{v}(\frac{1}{n(1-c)}+\frac{c}{(n-1)(1-c)^2})&
\tt{if} & 1<n \;
\end{array}
\end{equation}

where $\bar{v} := \int_c^1v\sigma(v)\D v = c +
\frac{n+1}{n+2}(1-c)$.

For $n=1$, we expect from the above formula that, since
$l[\sigma]<0$, we have the usual three phases seen in the pure
case, namely the \emph{low density}, the \emph{high density} and
the \emph{maximal current} phases.this expectation is borne out in
our simulation, where we have obtained the average density as a
function of $\alpha$, for various values of $\beta$ (Figure 5a).

As far as $c=\frac{1}{2}$, one can see after a simple calculation
that $l[\sigma]$ can never be positive. To obtain a positive $l$,
we take $n=2$ and $c=\frac{1}{3}$ (a broader spectrum of hopping
rates) for which $l$ is found to be equal to $3$. In this case
the results of simulations are shown in Figure 5b.

\begin{figure}[htb]
\begin{center}
\subfigure[$n=2,
c=\frac{1}{2}$]{\includegraphics[scale=0.55,viewport=0 0 500
390,clip]{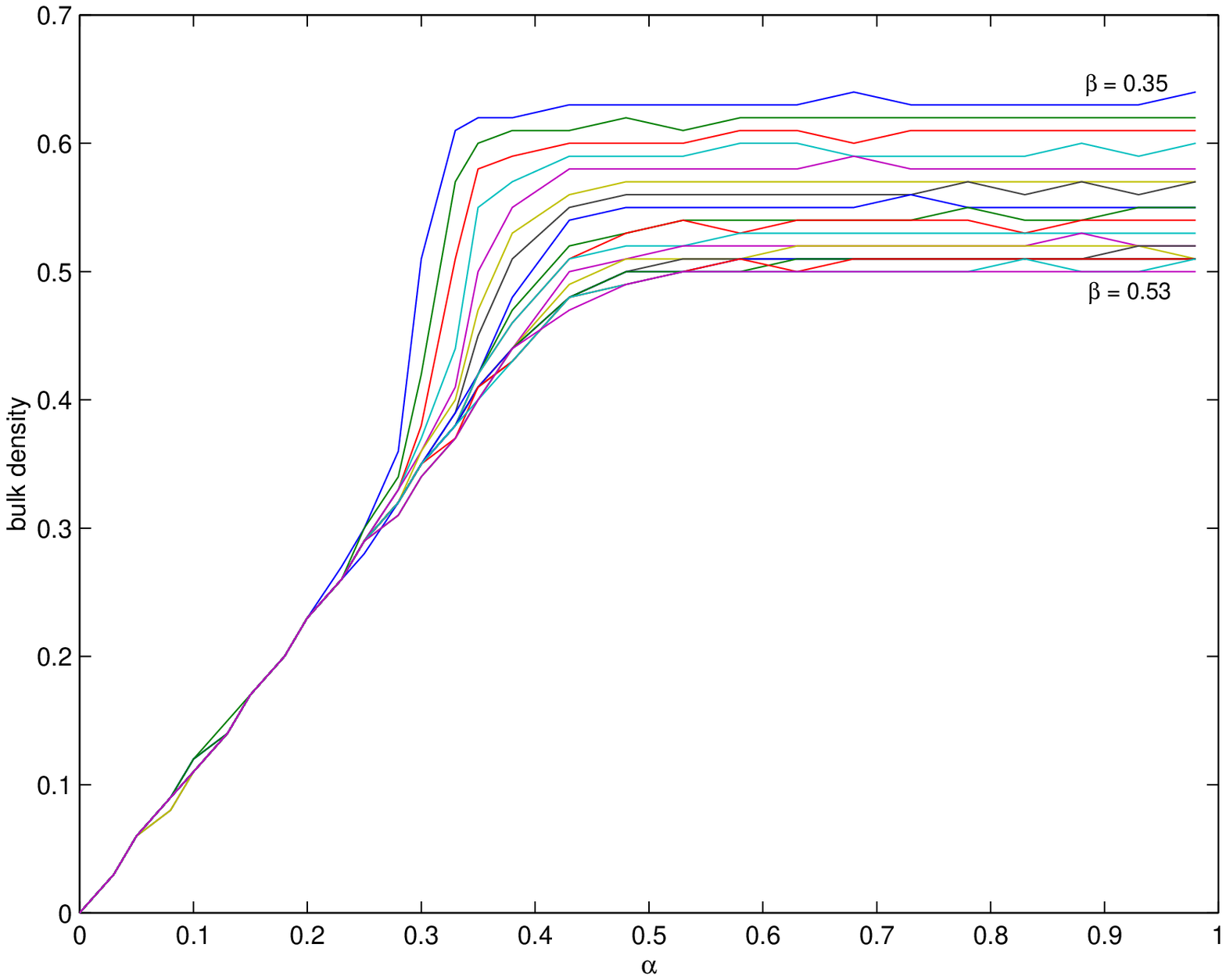}}%
\hspace{\subfigtopskip}%
\hspace{\subfigbottomskip} \subfigure[$n=2,
c=\frac{1}{3}$]{\includegraphics[scale=0.55,viewport=0 0 500
390,clip]{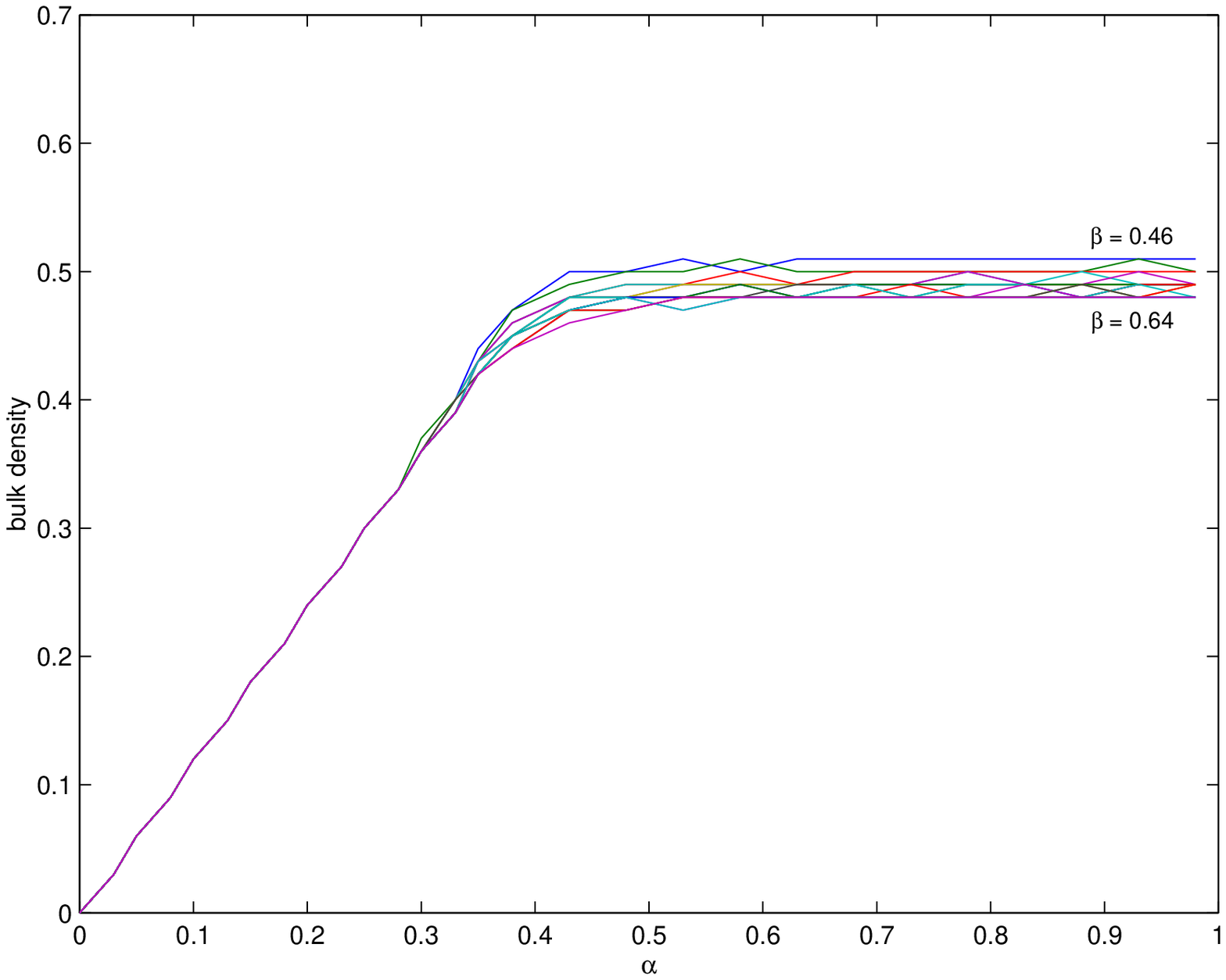}}
\end{center}
\caption{Bulk density versus $\alpha$ for the distribution
(12)($type~number = 200 $).}
\end{figure}

It is seen that when $l[\sigma]>0$,no jump discontinuity exists in
density, implying that there is no high density phase in the phase
diagram. For all $\beta$, there is a critical $\alpha_c$
independent of $\beta$, after which we enter the maximal current
phase ; this is in accord with the exact a analysis of \cite{K3}.
The corresponding phase diagram is shown in Figures 6a and 6b.

\begin{figure}[htb]
\begin{center}
\subfigure[$n=2,
c=\frac{1}{2}$]{\includegraphics[scale=0.55,viewport=0 0 500
390,clip]{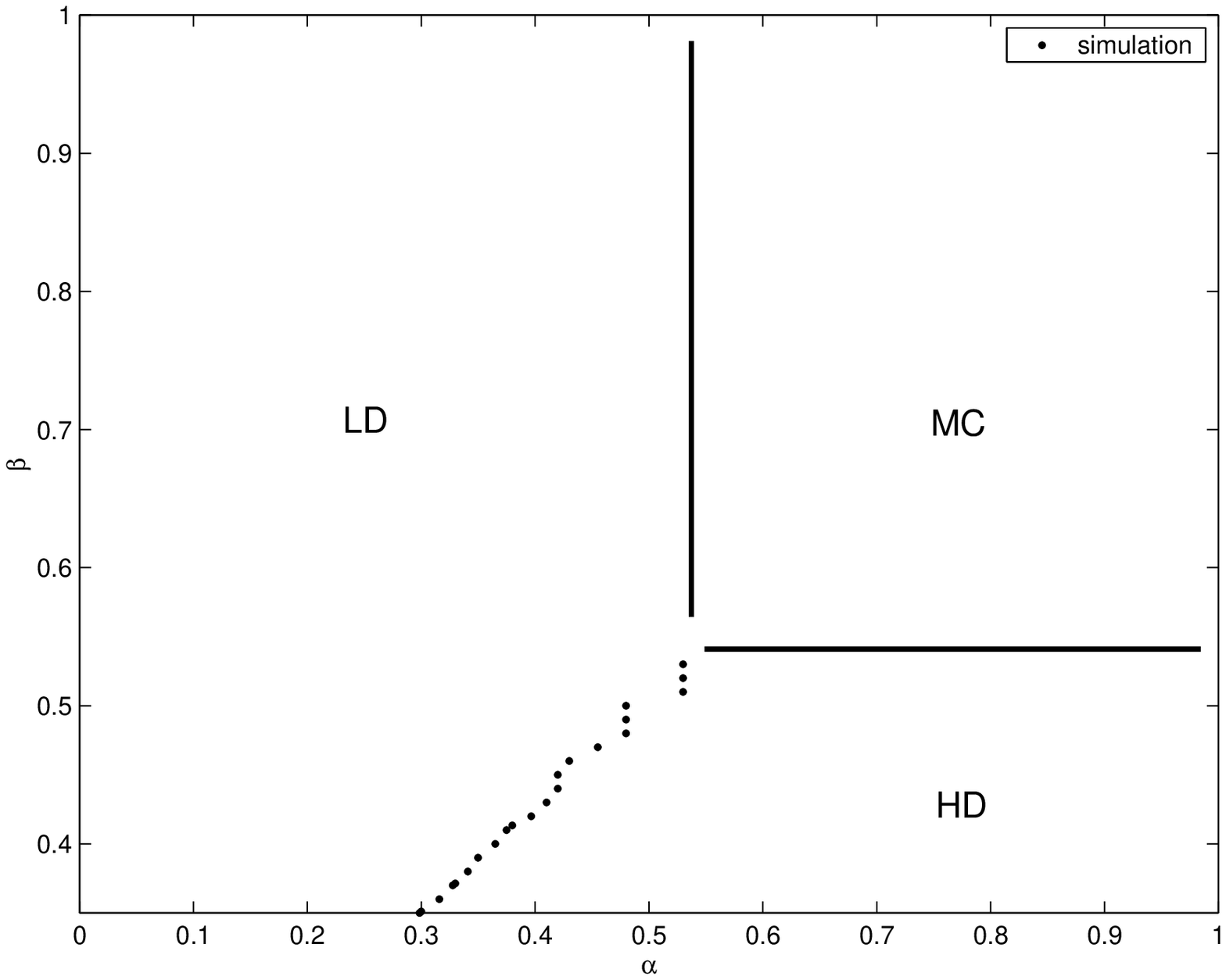}}%
\hspace{\subfigtopskip}%
\hspace{\subfigbottomskip} \subfigure[$n=2,
c=\frac{1}{3}$]{\includegraphics[scale=0.55,viewport=0 0 500
390,clip]{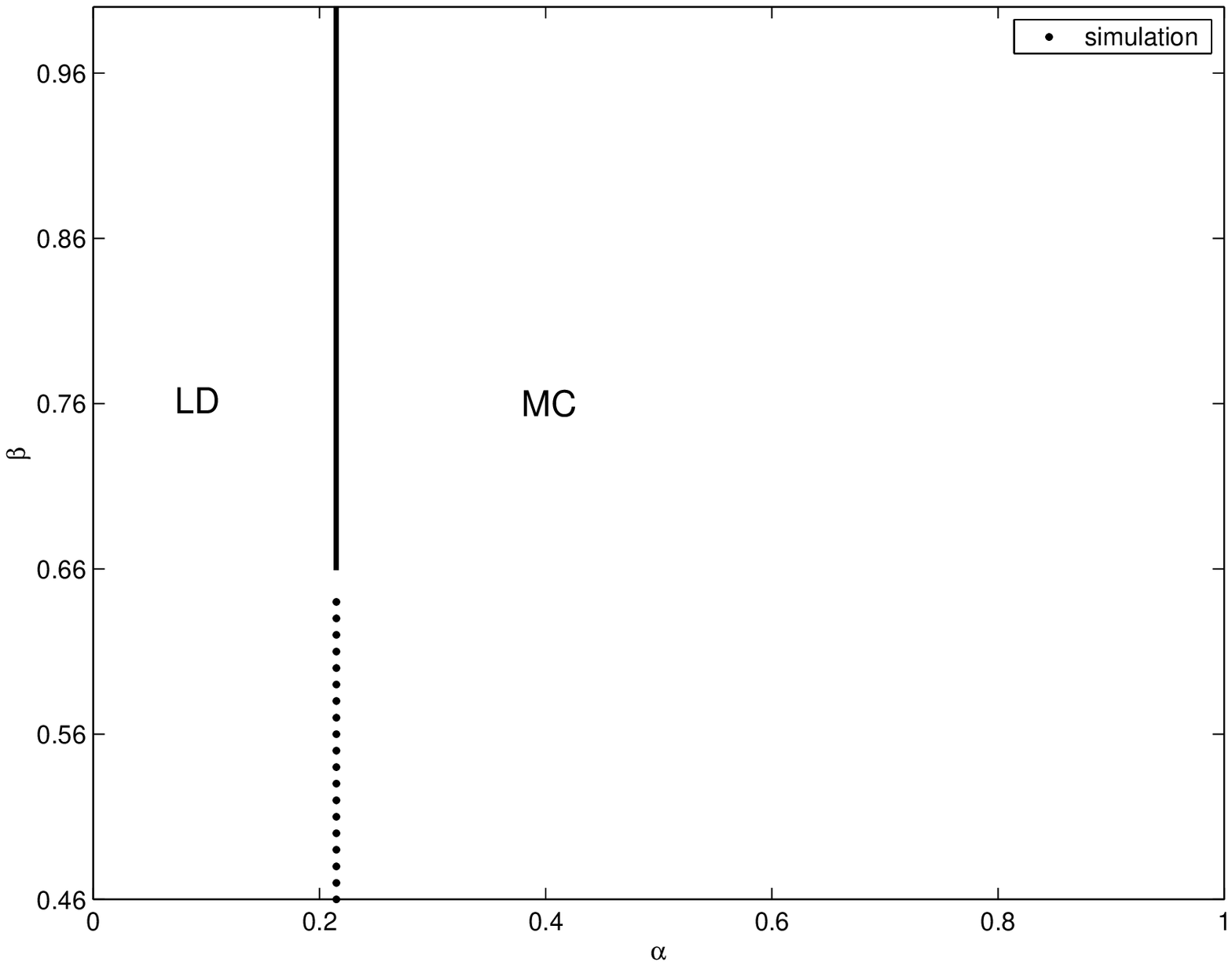}}
\end{center}
\caption{The phase diagram for problem \textbf{b} ($type~number =
200 $).}
\end{figure}

\subsection*{c) \begin{normalsize}The effect of a site impurity\end{normalsize}}
In this problem, our aim is to see what effect a site impurity
has on the density profiles of particles, when the particles have
the possibility of overtaking each other. For simplicity, we
consider a \emph{two} species model.

It is expected that such a blockage will mostly affect \emph{slow}
particles, since \emph{fast} particles have a chance of passing
by the particle accumulated behind the blockage. This is indeed
seen in our numerical solution. Figure 7a and 7b depict the
density profiles of both types of particles (\emph{fast} and
\emph{slow}) for two different values of $v_1$ and $v_2$. Density
profiles changes versus the strength of impurity $\lambda$ (the
\emph{slowing down factor}) is shown in Figure 10.

\begin{figure}[htb]
\begin{center}
\subfigure[$v_1=0.5,
v_2=1.5$]{\includegraphics[scale=0.9,viewport=0
0 320 240,clip]{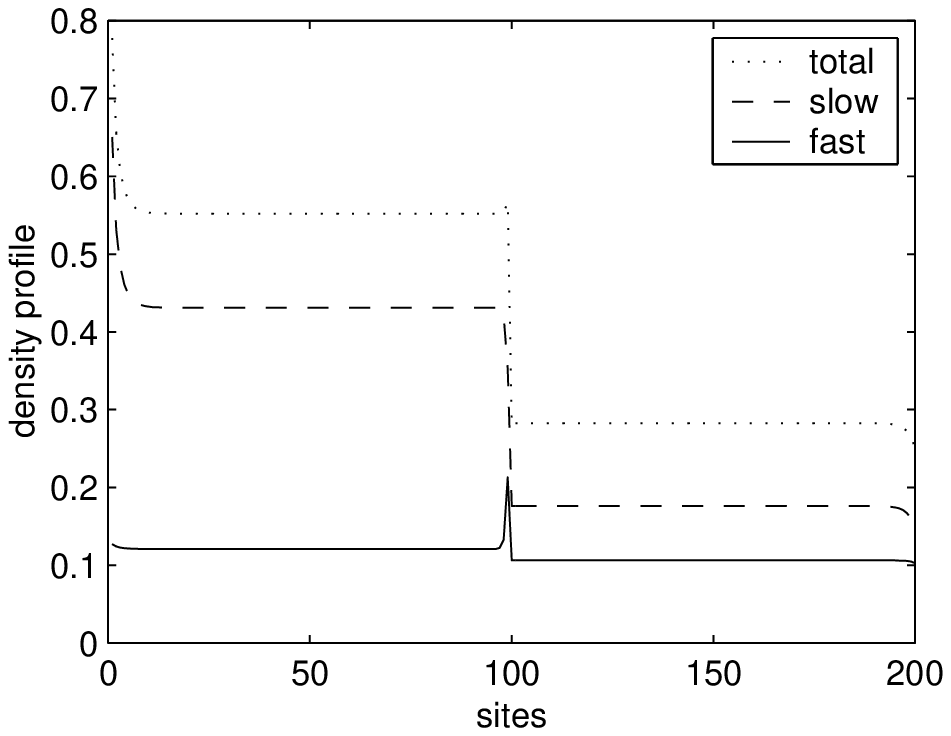}}%
\hspace{\subfigtopskip}%
\hspace{\subfigbottomskip} \subfigure[$v_1=0.8,
v_2=1.2$]{\includegraphics[scale=0.9,viewport=0 0 320
240,clip]{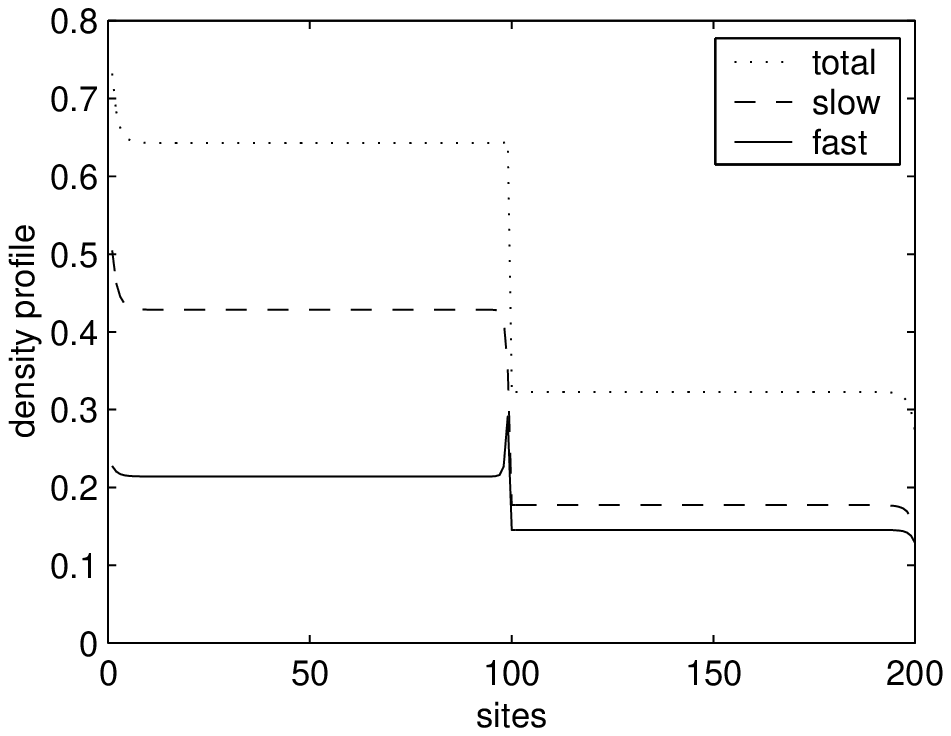}}
\end{center}
\caption{Density profile of \emph{slow} and \emph{fast} particles
in the presence of an impurity ($\lambda=0.5, \alpha=0.8,
\beta=0.8$). }
\end{figure}

\begin{figure}[t]
\begin{center}
\includegraphics[scale=0.7,viewport=0 0 350 240,clip]{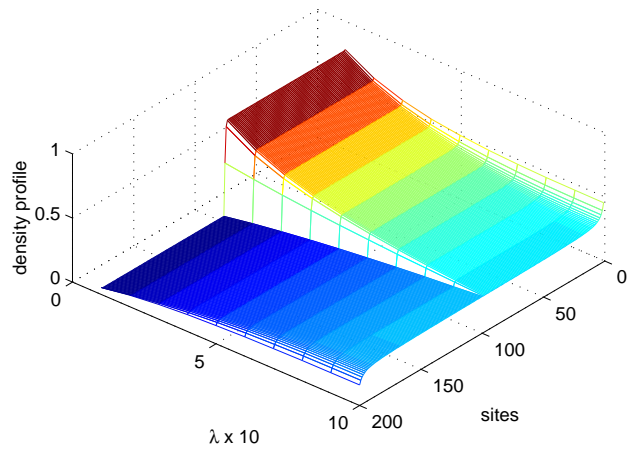}
\end{center}
\caption{Total density profile versus the strength of impurity
($v_1=0.8, v_2=1.2, \alpha=0.8, \beta=0.8$).}
\end{figure}

\section{Conclusion}
In this paper we have studied an asymmetric exclusion model on an
open chain with disorder in particle hopping rates. Following the
method of \cite{BB},and guided by the analysis of \cite{K3, K1},
we have shown how the distribution of hopping rates affects the
generic phase diagram of the single species ASEP. In particular
we have taken the hopping rates from the distribution

\begin{equation}
\sigma (v) = \frac{n+1}{(1-c)^{(n+1)}}(v-c)^n, c\leq v \leq 1
\end{equation}
also studied in \cite{BB}, and have shown that for $n=2$, the
high density phase may or may \emph{not} exist depending on the
value of $c$; i.e. \emph{the width of the distribution}. We have
also studied how in a single \emph{two} species process, with
hopping rates $v_1$ and $v_2$, a single site impurity affects the
density profiles of \emph{fast} and \emph{slow} particles.

\section*{Acknowledgments}

We would like to thank P~Abghari for providing hardware
facilities for simulations.


\end {document}